\documentclass[aps,pra,twocolumn,superscriptaddress,showpacs]{revtex4-1}
\usepackage{amsmath}
\usepackage{dcolumn}
\usepackage{amssymb}
\usepackage{amsfonts}
\usepackage{graphicx}
\usepackage[T1]{fontenc}
\usepackage{indentfirst}

\begin{document}
\title{Pfaffian-like ground states for bosonic atoms and molecules in one-dimensional optical lattices}
\author{Tanja \DJ uri\'c}
\affiliation{Instytut Fizyki im. M. Smoluchowskiego, Uniwersytet Jagiello\'nski, \L{}ojasiewicza 11, 30-348 Krak\'ow, Poland}
\author{Nicholas Chancellor}
\affiliation{London Centre for Nanotechnology, University College London, 17-19 Gordon Street, London, WC1H 0AH, United Kingdom}
\author{Philip J. D. Crowley}
\affiliation{London Centre for Nanotechnology, University College London, 17-19 Gordon Street, London, WC1H 0AH, United Kingdom}
\author{Pierfrancesco Di Cintio}
\affiliation{Universit\`a di Firenze and INFN, via G. Sansone 1, I-50019 Sesto Fiorentino, Italy}
\author{Andrew G. Green}
\affiliation{London Centre for Nanotechnology, University College London, 17-19 Gordon Street, London, WC1H 0AH, United Kingdom}

\date{\today}
\begin{abstract}
We study ground states and elementary excitations of a system of bosonic atoms and diatomic Feshbach molecules trapped in a one-dimensional optical lattice using exact diagonalization and 
variational Monte Carlo methods. We primarily study the case of an average filling of one boson per site. In agreement with bosonization theory, we show that 
the ground state of the system in the thermodynamic limit corresponds to the Pfaffian-like state when the system is tuned towards the superfluid-to-Mott insulator 
quantum phase transition. Our study clarifies the possibility of the creation of exotic Pfaffian-like states in realistic one-dimensional systems. 
We also present preliminary evidence that such states support non-Abelian anyonic excitations that have potential application for fault-tolerant topological quantum computation. 

\end{abstract} 
\maketitle
\section{Introduction}
\label{sec:intro}
\indent
The possibility of a fault-tolerant topological quantum computation \cite{Kitaev, Freedman,DasSarma,Nayak,Pachos,Pachos2} based upon topological quasiparticles that obey non-Abelian statistics (non-Abelian anyons) \cite{Moore,Stern1,Stern2} 
motivated much recent interest in the new systems that support such quasiparticles. The idea behind the topological quantum computation is that non-Abelian anyons could be 
used to encode and manipulate information in a way that is resistant to error. Namely, if a quantum system has topological degrees of freedom, like non-Abelian anyons, then the information
contained in those degrees of freedom will be protected against errors caused by local interactions with the environment. This provides the possibility of using such 
systems to perform fault-tolerant quantum computation without decoherence.

Non-Abelian states of matter also present the fundamental intellectual challenge of principle and of experimental realization \cite{DasSarma2, Stern3, Bonderson}. The understanding of the origin and properties of non-Abelian phases is far from complete and is at the frontier of current theoretical research.
The fundamental objectives are the understanding of the interplay between topology and quantum mechanics that leads to the formation of non-Abelian phases and the investigation of new models 
 that have non-Abelian quasiparticles \cite{Kitaev2,Teo}.

Non-Abelian anyons first appeared in the context of the fractional quantum Hall (FQH) effect \cite{Moore}, since FQH systems are believed to have a series of exotic 
non-Abelian states. Such states, like the Pfaffian state \cite{Greiter2,Nayak2,Read1,Read2,Wilkin,Cappelli}, which is the exact ground state of 
quantum Hall Hamiltonians with three-body contact interactions, have elementary excitations that are non-Abelian anyons. Similar states have also been predicted to occur 
in cold atoms \cite{Wilkin, Cooper, Gurarie,Moller,Duric2,Wu,Sterdyniak}, superconductors with $p$-wave pairing symmetry \cite{Nayak}, hybrid systems of superconductors with topological insulators 
and/or semiconductors \cite{Fu1,Fu2,Nilsson,Nikolic,Sau}, and non-Abelian lattice spin models \cite{Levin}.

 Although non-Abelian states are associated with two-dimensional (2D) systems, analogous states can be found in certain one-dimensional (1D) models \cite{Tsvelik1,Tsvelik2,Fendley,Tu,Paredes1,Paredes2,Nielsen, Greiter,Thomale}. 
 Ultimately, such 1D non-Abelian states must be braided in order to compute. This could be achieved by combining 1D systems into a 2D network as previously proposed in the case of
 Majorana fermions \cite{Alicea}. Understanding how to build states that support non-Abelian defects is an important building block towards topological computation. In this paper we 
 refine a previous proposal for a Pfaffian-like state that was proposed as an ansatz for the ground state of bosonic atoms subject to three-body infinite repulsive interactions and 
 in a 1D optical lattice \cite{Paredes1}. Although such three-body interactions are rare in nature, several experimentally realizable methods have been proposed to realize dominant three-body interactions between bosonic 
 atoms in optical lattices \cite{Paredes1,Cooper2,Will,Daley, Petrov1, Petrov2}. In particular, three-body interactions can be efficiently simulated by mixtures of bosonic atoms and molecules under conditions that are achievable with current technology 
 in systems of atoms and molecules in optical lattices \cite{Paredes1,Cooper2}.

 Therefore, the physical system that we consider is a collection of bosonic atoms and 
 diatomic Feshbach molecules trapped in a 1D optical lattice. Under certain experimentally achievable conditions, the system can be described by an effective Hamiltonian 
 for bosonic atoms with two-body and three-body contact interactions \cite{Paredes1}. We study the ground states and elementary excitations of the system in the limit of infinite repulsive three-body interactions and for 
 a range of values of the two-body interaction strength.

 The Pfaffian-like ansatz was originally proposed as an ansatz for the ground-state wave function of the 
 system in the absence of two-body interactions \cite{Paredes1}. However, our results show that the Pfaffian-like ansatz wave function most closely corresponds to the exact 
 ground-state wave function of the system at some finite value of the two-body interaction strength. The results also indicate that in the thermodynamic limit this value of the interaction strength 
 might be close to the value where the system undergoes a quantum phase transition from the superfluid state to the Mott insulating state, as previously found within the bosonization approach \cite{Lee}.

 Non-Abelian states of matter order their constituent particles following a hidden global pattern that is not associated with the breaking of 
 any symmetry \cite{Nayak,Wen}. This leads to a degeneracy that is not based upon simple symmetry considerations and is robust against perturbations and interactions 
 with the environment. Topological quasiparticles of such systems exhibit an exotic statistical behavior. Namely, the interchange of two identical quasiparticles 
 takes one ground state into another. If two different exchanges are performed consecutively among the quasiparticles, the final state of the system will depend upon the order in which these exchanges were 
 carried out. This ordering dependence is the reason why such states and their quasiparticles are called non-Abelian or noncommutative. In addition, 
 the quasiparticles of a non-Abelian system are neither fermions nor bosons, which motivated the name anyons.

 The Pfaffian-like states that we consider in this paper cannot be characterized by any local order parameter and exhibit a global hidden order that is associated with the organization of bosons in
 identical indistinguishable clusters. Indistinguishability between the clusters is achieved by symmetrization over the subsets of coordinates of each cluster. This symmetrization 
 introduces the possibility of topological degeneracy in the space of quasiparticles and makes these states potential carriers of non-Abelian excitations \cite{Paredes1,Paredes2,Paredes3}.

 Our explicit calculations use variationally optimized entangled-plaquette states for systems of up to 60 sites. 
 These are benchmarked with exact diagonalization (ED) studies of systems of up to 14 sites.

 Using the ED method, we first study ground-state properties and elementary excitations of the system for small system sizes and with periodic boundary conditions. 
 The Pfaffian-like ansatz was originally proposed as an ansatz for the ground-state wave function of the system in the absence of two-body interactions 
 (vanishing two-body interaction strength) \cite{Paredes1}. However, our ED results clearly demonstrate that the Pfaffian-like ansatz better approximates the ground state of the system at some finite value of the two-body interaction 
 strength. This interaction strength increases with increasing system size. Also, the overlap of the exact ground-state wave function at such a value of the two-body interaction strength and the Pfaffian-like ansatz wave function, decreases more gradually 
 with increasing system size in the presence of two-body repulsion than it does in the absence of the two-body interactions.

 The ED results thus indicate that in the thermodynamic limit the Pfaffian-like ansatz wave function most closely corresponds to the exact ground-state wave function of the system at 
 some finite value of the two-body interaction strength. This might be close to the value where the system undergoes a quantum phase transition from the superfluid state 
 to the Mott insulating state, as previously found within the bosonization approach \cite{Lee}. We also present preliminary evidence that these states support non-Abelian excitations required 
 for topological quantum computation.

 We further study the ground-state properties of the system for larger system sizes. Motivated by the recent success of tensor network methods \cite{Orus} to numerically simulate a variety of strongly correlated models, we use the entangled-plaquette-state (EPS) ansatz,
 also called the correlator-product-state (CPS) ansatz, and the variational Monte Carlo (VMC) method \cite{Changlani,Mezzacapo1,Neuscamman1,Mezzacapo2,Mezzacapo3,Mezzacapo4,Mezzacapo5,Neuscamman2,AlAssam,Duric}.
  In the EPS approach, the lattice is covered with overlapping plaquettes and the ground-state wave function is written in terms of the plaquette coefficients. Configurational weights are 
  then optimized using a VMC algorithm. Here, the plaquette coefficients that minimize the energy are found using the stochastic minimization method \cite{AlAssam,Duric,Sandvik,Lou}. For small 
  system sizes we find that the EPS and VMC calculation gives quite accurate estimates of the ground-state energy and the one-body and two-body correlation functions.

 To examine the proximity of the ground state to the Pfaffian-like ansatz for larger system sizes, we calculate the one-body and two-body correlation functions 
 for the exact ground-state and the Pfaffian-like ansatz wave functions and compare their asymptotic behavior. Since the EPS wave function gives quite accurate estimates 
 of the correlations within any plaquette, we estimate the asymptotic behavior of the correlation functions from the values of the correlation functions for the lattice 
 sites within a plaquette.

 We study the ground-state properties of the system for the system sizes $L=40$ and 60 sites. The results obtained within the EPS and VMC approach are consistent 
 with the ED results for smaller system sizes and with the results for vanishing two-body interaction strength obtained previously using variational matrix product 
 states (MPSs). For the system size $L=60$ sites, the maximum system size that we have considered, the results indicate that at some finite value of the two-body interaction strength $U/t=\bar{U}_C(L)$ 
 the exact ground-state wave function is still very close to the Pfaffian-like ansatz wave function.

The paper is organized as follows. In Sec. II we introduce
the effective three-body interacting atomic Hamiltonian for a system of bosonic atoms and diatomic Feshbach molecules trapped in a 1D optical lattice.
In Sec. III we review the theory of the Pfaffian-like states in 1D. In Sec. IV we present ED results
for small system sizes and with periodic boundary conditions. The results
for larger system sizes obtained within the EPS and VMC
approach are presented in Sec. V. In the final section, Sec. VI,
we draw our conclusions and discuss possible directions for
future research.

\section{Effective three-body interacting atomic Hamiltonian}
\label{sec:Hamiltonian}
We consider a systems of bosonic atoms and diatomic Feshbach molecules trapped in a 1D optical lattice. The system can be described by the Hamiltonian \cite{Paredes1,Timmermans,Holland}
\begin{equation}\label{eq:Hamiltonian1}
H = H_K + H_F + H_I ,
\end{equation}
where
\begin{equation}\label{eq:Hamiltonian2_1}
H_K = -t_a\sum_i(a_i^\dagger a_{i+1} + h.c.)-t_m\sum_i(m_i^\dagger m_{i+1} + h.c.), \nonumber
\end{equation}
\begin{equation}\label{eq:Hamiltonian2_2}
H_F = \sum_i[\delta m_i^\dagger m_i + \frac{U_{aa}}{2}a_i^\dagger a_i^\dagger a_ia_i + \frac{g}{\sqrt{2}} (m_i^\dagger a_ia_i + h.c.)], \nonumber
\end{equation}
and
\begin{equation}\label{eq:Hamiltonian2_3}
H_I = U_{am}\sum_i m_i^\dagger a_i ^\dagger a_i m_i +\frac{U_{mm}}{2}\sum_i m_i^ \dagger m_i^\dagger m_i m_i.\nonumber
\end{equation}
The bosonic operators for atoms and molecules are denoted $a_i$ and $m_i$, respectively. The term $H_K$ describes the tunneling processes 
of atoms and molecules. The term $H_F$ is the Feshbach resonance term 
and the term $H_I$ describes the on-site atom-molecule and molecule-molecule interactions. Here $t_a$, $t_m$, $U_{aa}$, $U_{am}$, and $U_{mm}$ are hopping matrix elements
for atoms and molecules and the on-site atom-atom, atom-molecule, and molecule-molecule interaction strengths, respectively. The energy offset between open
and closed channels in the Feshbach resonance model is denoted $\delta$, and $g$ is the coupling strength to the closed channel.

We further assume $U_{aa}$, $U_{am}$, $U_{mm} \geq 0$, and $\delta > 0$. In the limit $\gamma^2=g^2/2\delta^2 \ll 1$, the formation of molecules is highly suppressed and 
the effective Hamiltonian for the system can be obtained, to first order in $\gamma^2$, by projection of the Hamiltonian, (\ref{eq:Hamiltonian1}), onto the subspace with no molecules. 
The resulting effective Hamiltonian is \cite{Paredes1}
\begin{eqnarray}\label{eq:Hamiltonian3}
H_{eff}&=&-t_a\sum_i(a_i^\dagger a_i+h.c.)+ U_{am}\gamma^2 \sum_i(a_i^\dagger)^3(a_i)^3 \nonumber \\
&-&t_m\gamma^2\sum_i \left[(a_i^\dagger)^2(a_{i+1})^2+h.c.\right] \\
&+&(U_{aa}-g^2/\delta)\sum_i(a_i^\dagger)^2(a_i)^2. \nonumber
\end{eqnarray}
In the limit $t_m\gamma^2 \ll t_a$, valid in typical experiments with $^{87}$Rb \cite{Paredes1}, the effective Hamiltonian, (\ref{eq:Hamiltonian3}), further reduces to
\begin{eqnarray}\label{eq:Hamiltonian4}
H_{eff} &=& -t\sum_i(a_i^\dagger a_{i+1}+h.c.)+U_2\sum_i (a_i^\dagger)^2 (a_i)^2 \nonumber\\
&+&U_3\sum_i(a_i^\dagger)^3(a_i)^3, 
\end{eqnarray}
with $t=t_a$, $U_2=U_{aa}-g^2/\delta$, and $U_3=U_{am}\gamma^2$. The effective Hamiltonian, (\ref{eq:Hamiltonian4}), is the Hamiltonian for a system of bosonic atoms in a 1D optical 
lattice with repulsive two- and three-body on-site interactions.

We further assume that $U_3 \gg t_a$: the limit accessible in typical setups with $^{87}$Rb atoms \cite{Paredes1}.
In the limit $U_3\rightarrow \infty$, the Hilbert space is projected onto the subspace of states with occupation numbers $n_i = 0,1,2$. The bosonic operators subject to this 
condition, that is, the condition $(a_{3,i}^\dagger)^3=0$, are referred to as three-hard-core bosonic operators and satisfy the commutation relations
$[a_{3,i},a_{3,j}^\dagger]=\delta_{i,j}(1-\frac{3}{2}(a^\dagger _{3,i})^2(a_{3,i})^2)$. Since $a_{3,i}^\dagger|n_i\rangle=(1-\delta_{n_i,2})\sqrt{n_i+1}|n_i+1\rangle$, these operators can be represented by $3\times 3$ matrices of the form
\begin{equation}\label{eq:a3}
a_{3,i} = \left(\begin{array}{ccc}
0 & 1 & 0 \\
0 & 0 & \sqrt{2} \\
0 & 0 & 0 \end{array} \right).
\end{equation}
In terms of these three-hard-core bosons the projected effective Hamiltonian is 
\begin{equation}\label{eq:Hamiltonian5}
H_{eff} = -t\sum_i (a_{3,i}^\dagger a_{3,i+1}+h.c.) + \frac{U}{2}\sum_i(a_{3,i}^\dagger)^2(a_{3,i})^2, 
\end{equation}
with $U=2U_2$. We also note that within the experimental situation that we study, three-body losses are strongly suppressed (the binding energy, and therefore the released energy, is not larger 
than the lattice depth) \cite{Paredes1}. In this paper we study the ground states of the Hamiltonian, (\ref{eq:Hamiltonian5}), for a range of values of $U/t$ and at the fixed average filling factor of one 
boson per lattice site. We compare the ground states to the Pfaffian-like ground-state ansatz wave function \cite{Paredes1}. The properties
of the Pfaffian-like state are reviewed in the following section.

\section{Pfaffian-like states} 
\label{sec:Pfaffian}
The idea of using symmetrized indistinguishable cluster states as ansatz wave functions for non-Abelian 1D bosonic liquids was proposed by
Paredes, Kielmann, and Cirac \cite{Paredes1,Paredes2,Paredes3}. The Pfaffian-like ansatz, inspired by the form of the ground state for fractional quantum Hall
bosons subject to a three-body interaction \cite{Read1,Read2,Wilkin}, was originally proposed as an ansatz for the ground-state wave function of the Hamiltonian, (\ref{eq:Hamiltonian5}),
at $U=0$ \cite{Paredes1}.

For bosons in the lowest Landau level subject to the three-body interaction potential $U_3\sum_{i\neq j\neq k} \delta^2(z_i-z_j)\delta^2(z_i-z_k)$, with $z_i=x_i+iy_i$ being
the complex coordinate in the 2D plane, the exact ground state in the limit $U_3\rightarrow \infty$ is the Pfaffian state \cite{Wilkin,Cappelli,Greiter,Thomale}
\begin{equation}\label{eq:Pfaffian2D}
\Phi_{3} \propto \emph{S}_{\uparrow,\downarrow} \{ \prod_{i<j}^{N/2}(z_i^\uparrow-z_j^\uparrow)^2 \prod_{i<j}^{N/2} (z_i^\downarrow-z_j^\downarrow)^2 \}.
\end{equation}
This state is a symmetrized product of two identical Laughlin states \cite{Laughlin},
\begin{equation}\label{eq:Laughlin}
\Phi_2^\sigma \propto \sum_{i<j}^{N/2}(z_i^\sigma-z_j^\sigma)^2, 
\end{equation}
with $\sigma = \uparrow, \downarrow$. Since the Laughlin state of each cluster is a zero-energy eigenstate of the two-body interaction potential $\sum_{i\neq j}\delta(z_i-z_j)$, 
 three particles can never coincide in a state of the form of (\ref{eq:Pfaffian2D}).  The operator that symmetrizes over the two virtual subsets of coordinates 
$\{z_i^\uparrow\}$ and $\{z_i^\downarrow\}$ is denoted $\emph{S}_{\uparrow,\downarrow}$.

\begin{figure}[t!]
\includegraphics[width=\columnwidth]{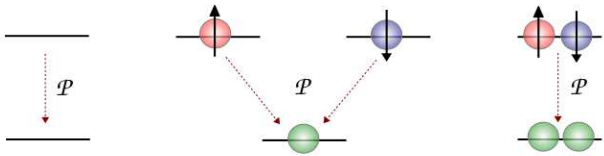}
 \caption{\label{fig:P} Schematic of the local projector $P_i$ at a lattice site $i$. The operator $P_i$ projects the two identical local degrees
 of freedom onto a new degree of freedom that is symmetric under exchange of the two components. Here the operator $P_i$ maps the single-site four-dimensional 
 Hilbert space of two species of hard-core bosons, $\uparrow$ and $\downarrow$ (red and blue spheres) to the single-site three-dimensional Hilbert space of three-hard-core-bosons (green spheres).   } 
 \end{figure}
An ansatz for the ground state of the Hamiltonian, (\ref{eq:Hamiltonian5}), at $U=0$ was proposed \cite{Paredes1} in direct analogy with the wave function, (\ref{eq:Pfaffian2D}):
\begin{equation}\label{eq:Pfaffian1D}
\Psi_3 \propto \emph{S}_{\uparrow,\downarrow} \{ \prod_{i<j}^{N/2}|\mbox{sin}(x_i^\uparrow-x_j^\uparrow)| \prod_{i<j}^{N/2} |\mbox{sin}(x_i^\downarrow-x_j^\downarrow)|\}.
\end{equation}
This ansatz has the same form as the Pfaffian state, (\ref{eq:Pfaffian2D}), with the Laughlin state replaced by a Tonks-Girardeau state \cite{Girardeau}: 
\begin{equation}\label{eq:TGstate}
\Psi_2^\sigma \propto\prod_{i<j}^{N/2} |\mbox{sin}(x_i^\sigma-x_j^\sigma)|. 
\end{equation}
The Tonks-Girardeau state, (\ref{eq:TGstate}), is the ground state of 1D lattice hard-core bosons described by the Hamiltonian 
$H_{2,\sigma} = -t \sum_i (a_{2,\sigma,i}^\dagger a_{2,\sigma,i+1} +h.c.)$ and with periodic boundary conditions \cite{Paredes4}. The hard-core bosonic operators $a_{2,\sigma,i}$ obey $(a_{2,\sigma,i}^\dagger)^2=0$, 
allowing only occupation numbers of $n_i^\sigma=0$ or 1 boson per site. Here $x_i^\sigma = 2\pi/Li$ and $i=1,...,L$, with $L$ being the number of lattice sites.

To write the ansatz wave function in second quantized form, we define a projection operator $\mathcal{P}$ such that 
\begin{equation}\label{eq:Pfaffian1D_2}
|\Psi_3\rangle = \mathcal{P}(|\Psi_2^\uparrow\rangle \otimes |\Psi_2^\downarrow\rangle).
\end{equation}
The projection operator $\mathcal{P}$ is a local operator of the form 
 \begin{equation}\label{eq:P}
 \mathcal{P}=\mathcal{P}_i^{\otimes L}, 
 \end{equation}
where $L$ is the number of lattice sites and $\mathcal{P}_i$ is the local projector at a lattice site $i$,
 \begin{equation}\label{eq:Pl}
 \mathcal{P}_i= \left(\begin{array}{cccc}
1 & 0 & 0 & 0 \\
0 & 1 & 1 & 0 \\
0 & 0 & 0 & \sqrt{2} \end{array} \right).
 \end{equation}
 $P_i$ maps the single site four-dimensional Hilbert space of two species of hard-core bosons, 
 $\uparrow$ and $\downarrow$, to the single-site three-dimensional Hilbert space of three-hard-core bosons as illustrated in Fig. 1.

 It can be further shown that the one- and two-body correlation functions for the ansatz wave function, (\ref{eq:Pfaffian1D_2}), have the asymptotic behavior \cite{Paredes1,Paredes4}
 \begin{eqnarray}
 \langle a_{i+\Delta}^\dagger a_i\rangle\rightarrow \Delta^{-1/4},\\
 \langle a_{i+\Delta}^\dagger a_{i+\Delta}^\dagger a_i a_i\rangle\rightarrow \Delta^{-1},\nonumber
 \end{eqnarray}
 for large $\Delta$ and for a large system size $L$. The two-body correlation function corresponds to the one-particle correlation function for on-site pairs \cite{Paredes1},
 \begin{eqnarray}
  &&\langle a_{i+\Delta}^\dagger a_{i+\Delta}^\dagger a_i a_i\rangle\propto \\
  &&\langle \Psi_2^\uparrow|a^\dagger_{2,\uparrow,i+\Delta}a_{2,\uparrow,i}|\Psi_2^\uparrow\rangle\langle \Psi_2^\downarrow|a^\dagger_{2,\downarrow,i+\Delta}a_{2,\downarrow,i}|\Psi_2^\downarrow\rangle\nonumber\\
  &&\rightarrow \Delta^{-1/2}\Delta^{-1/2}\nonumber,
 \end{eqnarray}
 where $\langle \Psi_2^\sigma|a^\dagger_{2,\sigma,i+\Delta}a_{2,\sigma,i}|\Psi_2^\sigma\rangle\rightarrow \Delta^{-1/2}$, for large $\Delta$, is the well known result for a Tonks-Girardeau gas \cite{Paredes4}. 
 In other words, although the system of atoms has some kind of coherence, with slowly decaying spatial correlations $\propto \Delta^{-1/4}$, the underlying system of on-site
 pairs is in a much more disordered state with a fast decay of spatial correlations $\propto \Delta^{-1}$.
 
 \section{Exact diagonalization results for small system sizes}
\label{sec:ED}
\begin{figure}[b!]
\includegraphics[width=\columnwidth]{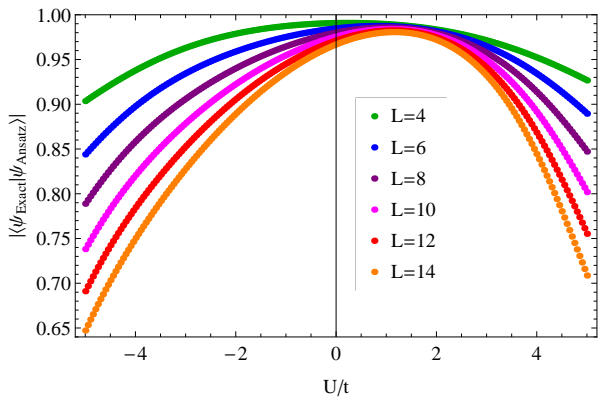}
 \caption{\label{fig:Overlap} Overlap of the exact ground-state wave function of the Hamiltonian, (\ref{eq:Hamiltonian5}), and the ansatz wave function, (\ref{eq:Pfaffian1D_2}),
 as a function of the two-body interaction strength $U/t$ for system sizes of $L\leq 14$ sites. Here the filling factor $\nu=N/L =1$, with $N$ being the number of particles.} 
 \end{figure}
We first examine the ground-state properties and elementary excitations of the system for small system sizes and with periodic boundary conditions using the ED method. The ground-state properties are calculated 
for a range of values of the two-body interaction strength $U/t$ and at the filling factor of one particle per site. We calculate the overlap of the exact ground-state
wave function of the Hamiltonian, (\ref{eq:Hamiltonian5}), and the ansatz wave function, (\ref{eq:Pfaffian1D_2}), 
the average occupation of sites with one and two particles, and the one- and two-body correlation functions.
\\
\indent 
 \begin{figure}[t!]
 \includegraphics[width=\columnwidth]{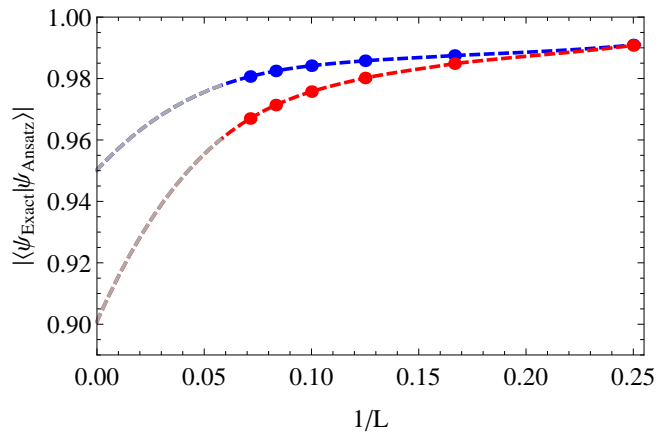}
 \caption{\label{fig:Overlap0Uc}Overlap of the exact ground-state wave function of the Hamiltonian, (\ref{eq:Hamiltonian5}), and the ansatz wave function, (\ref{eq:Pfaffian1D_2}), 
 at the filling factor $\nu=1$ for the values of the two-body interaction strength $U/t=0$ (red symbols) and $U/t=\bar{U}_C(L)$, where 
 the overlap $|\langle\psi_{Exact}|\psi_{Ansatz}\rangle|$ is maximal (blue symbols). 
 Dashed gray lines are included as guides for the eye and were obtained by the extrapolation of the ED results for $L\leq 14$. 
 } 
 \end{figure}
 \begin{figure}[b!]
 \includegraphics[width=\columnwidth]{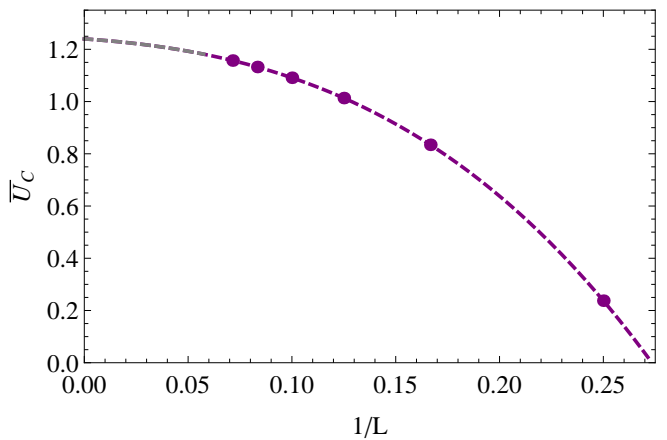}
 \caption{\label{fig:Uc}The value of the two-body interaction strength $U/t$ where the overlap $|\langle\psi_{Exact}|\psi_{Ansatz}\rangle|$ is maximal ($\bar{U}_C$) for 
 system sizes of $L\leq 14$ sites and at filling factor $\nu=1$. Dashed gray lines are included as guides for the eye and were obtained by the extrapolation of the ED results for $L\leq 14$.
 } 
 \end{figure}
 \begin{figure}[t!]
 \includegraphics[width=\columnwidth]{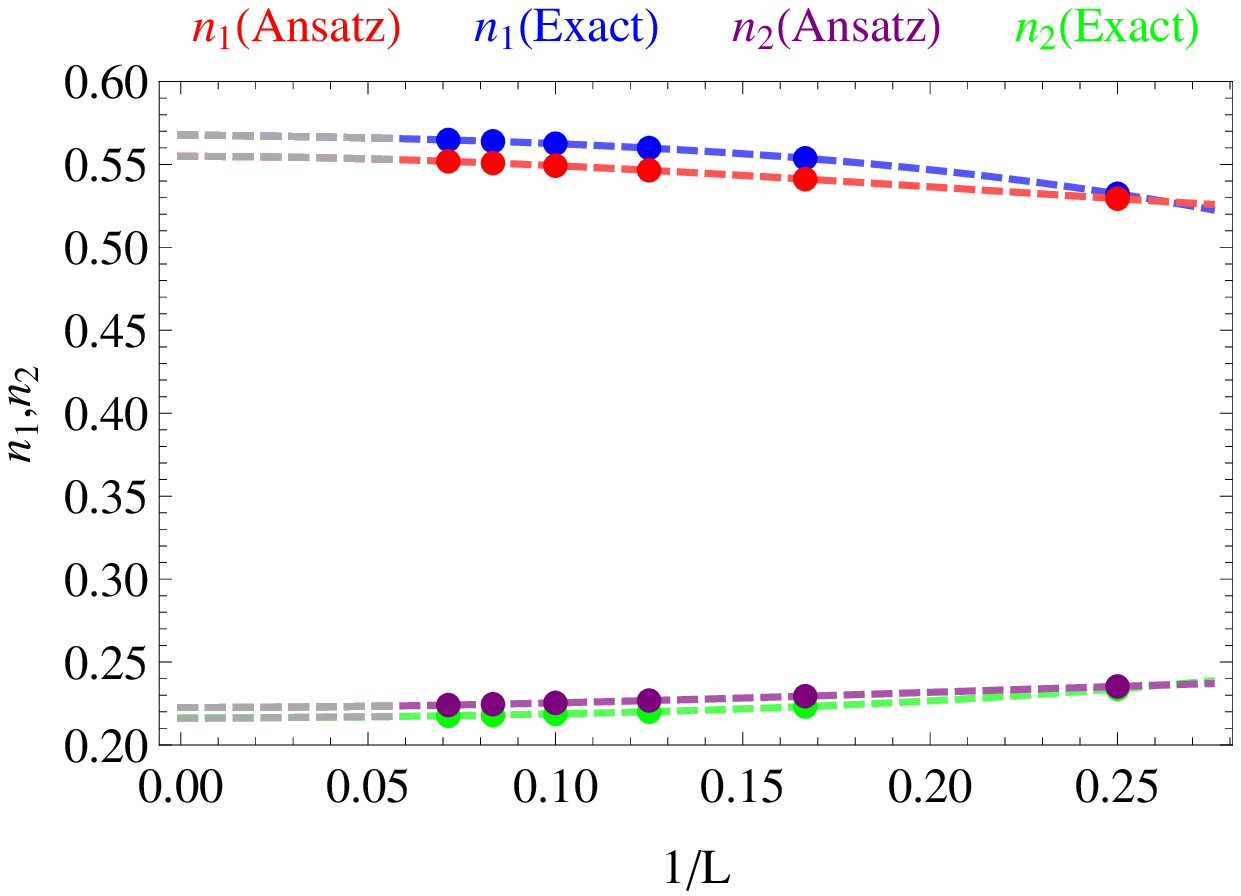}
 \caption{\label{fig:n1n2}
 Average number of sites with one particle $n_1=\frac{1}{L}\sum_{i=1}^L\langle n_i(2-n_i)\rangle$ (blue and red symbols) and two 
 particles $n_2=\frac{1}{2L}\sum_{i=1}^L\langle a_i^\dagger a_i^\dagger a_ia_i\rangle$ (green and purple symbols) for the 
 exact ground-state wave function of the Hamiltonian, (\ref{eq:Hamiltonian5}), at $U/t=\bar{U}_C(L)$ (blue and green symbols) and the ansatz wave function, 
 (\ref{eq:Pfaffian1D_2}), (red and purple symbols). Here the filling factor $\nu=1$ and $\bar{U}_C(L)$ denotes the value of the two-body interaction strength where 
 the overlap $|\langle\psi_{Exact}|\psi_{Ansatz}\rangle|$ is maximal. Dashed gray lines are included as guides for the eye and were obtained by the extrapolation of 
 the ED results for $L\leq 14$.
} 
 \end{figure}
The overlap of the exact ground-state wave function and the Pfaffian-like ansatz wave function for the system sizes $L\leq 14$ and as a function of the two-body interaction strength
$\bar{U}=U/t$ is shown in Fig. \ref{fig:Overlap}. The results clearly demonstrate that the Pfaffian-like ansatz wave function, (\ref{eq:Pfaffian1D_2}), is a better ansatz for the exact ground-state wave function of the Hamiltonian, 
(\ref{eq:Hamiltonian5}), at some finite value of the two-body interaction strength $\bar{U}_C(L)$ than it is for the exact ground-state wave function at $U=0$ as suggested previously \cite{Paredes1}.

Also, the overlap decreases more gradually with increasing system size $L$ at $\bar{U}_C(L)$ than it decreases at $U=0$ (Fig. \ref{fig:Overlap0Uc}). This indicates that in the 
thermodynamic limit ($L\rightarrow\infty$), the Pfaffian-like ansatz wave function most closely corresponds to the exact ground-state wave function of the Hamiltonian, (\ref{eq:Hamiltonian5}),
at some finite value of the two-body interaction strength $\bar{U}_C(L)$.

\begin{figure}[t!]
\includegraphics[width=\columnwidth]{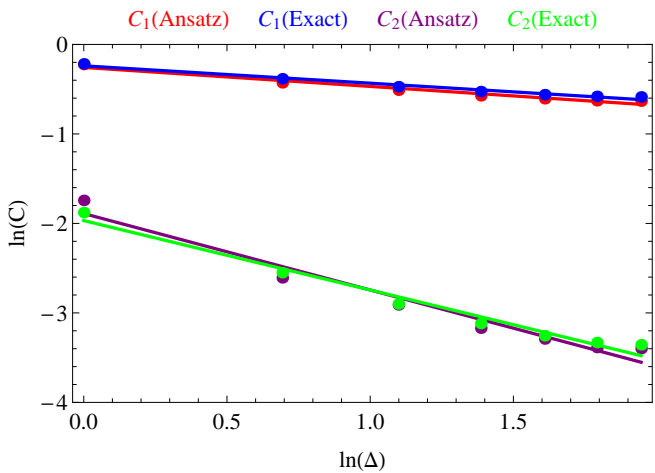}
 \caption{\label{fig:lnC} The one-body (blue and red symbols) and two-body (green and purple symbols) correlation functions for the exact ground-state wave function of the Hamiltonian, 
 (\ref{eq:Hamiltonian5}), at $U/t=\bar{U}_C(L)$ (blue and green symbols) and the ansatz wave function, (\ref{eq:Pfaffian1D_2}) (red and purple symbols), for the system size $L=14$. 
 Here the filling factor $\nu=1$ and $\bar{U}_C(L)$ denotes the value of the two-body interaction strength where the overlap 
 $|\langle\psi_{Exact}|\psi_{Ansatz}\rangle|$ is maximal.
 The long-distance scaling of the correlation functions $C_1(\Delta)\propto \Delta^{-\alpha_1}$ and $C_2(\Delta)\propto 
 \Delta^{-\alpha_2}$ is $\alpha_1\approx 0.194$ and $\alpha_2\approx 0.776$ for the exact ground state and $\alpha_1\approx 0.212$ and $\alpha_2\approx 0.854$ for the ansatz wave function.} 
 \end{figure}
The ED results also indicate that the value of $\bar{U}_C$ where the overlap is maximal increases with increasing system size (Fig. \ref{fig:Uc}).
It can also be shown that the value of $\bar{U}$ where the system undergoes a quantum phase transition from the superfluid state to the Mott insulating state, $\bar{U}_{SF-MI}$, decreases
with increasing system size. The ED results thus suggest that the value of $\bar{U}_C$ approaches the value of $\bar{U}_{SF-MI}$ with an increase in the system size $L$
and that the Pfaffian-like state might be the state at the superfluid-to-Mott insulator boundary as previously found within the bosonization approach \cite{Lee}.

We have further calculated the average number of sites with one particle, $n_1$, and with two particles, $n_2$, at the filling factor $\nu=1$ and for a range of values 
of the two-body interaction strength $\bar{U}=U/t$. The average number of sites with one particle and with two particles is
\begin{equation}\label{eq:n1n2}
n_1=\frac{1}{L}\sum_{i=1}^L\langle n_i(2-n_i)\rangle, 
\end{equation}
\begin{equation}
n_2=\frac{1}{2L}\sum_{i=1}^L\langle a_i^\dagger a_i^\dagger a_ia_i\rangle, \nonumber
\end{equation}
where $n_i=a_i^\dagger a_i$ and $n_1+2n_2=\nu=1$. 
The values of $n_1$ and $n_2$ for the Pfaffian-like ansatz wave function, (\ref{eq:Pfaffian1D_2}), and for the exact ground-state wave function of the Hamiltonian, 
(\ref{eq:Hamiltonian5}), at $U/t=\bar{U}_C(L)$ (where the overlap $|\langle\psi_{Exact}|\psi_{Ansatz}\rangle|$ is maximal) are shown in Fig. \ref{fig:n1n2}.
The values of $n_1$ and $n_2$ for the exact ground-state wave function are very close to the values of $n_1$ and $n_2$ for the Pfaffian-like ansatz wave function, as can
be clearly seen in Fig. \ref{fig:n1n2}.

We have also calculated the one-body and two-body correlation functions,
 \begin{eqnarray}\label{eq:C1C2}
 C_1&=&\langle a_{i+\Delta}^\dagger a_i\rangle,\\
 C_2&=&\langle a_{i+\Delta}^\dagger a_{i+\Delta}^\dagger a_i a_i\rangle, \nonumber
 \end{eqnarray}
for the exact ground-state wave function at $U/t=\bar{U}_C(L)$ and for the ansatz wave function for a system of $L=14$ sites. The results in Fig. \ref{fig:lnC} show that the
correlation functions for the exact and ansatz wave functions show very similar asymptotic behavior.

\begin{figure}[t!]
\includegraphics[width=\columnwidth]{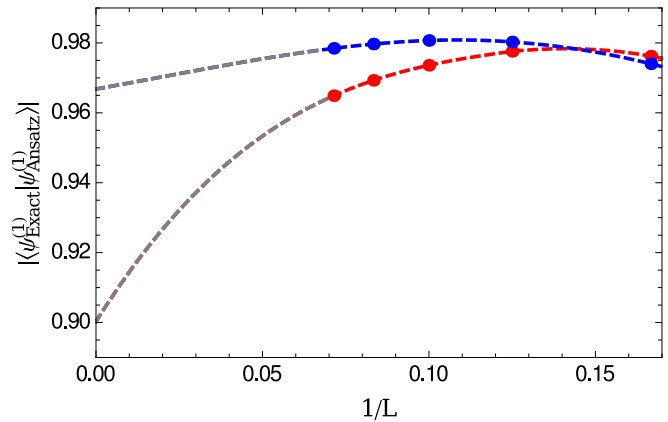}
\caption{\label{fig:Overlap0Uc_es}Overlap $O_1=O_2\equiv O$, (\ref{eq:ovl_es}), for two degenerate first excited states of the Hamiltonian, (\ref{eq:Hamiltonian5}), at the filling factor $\nu=1$ for values of the two-body interaction strength $U/t=0$ (red symbols) and $U/t=\bar{U}_C(L)$, where 
the overlap $|\langle\psi_{Exact}|\psi_{Ansatz}\rangle|$ for the ground state of the Hamiltonian, (\ref{eq:Hamiltonian5}), is maximal (blue symbols). 
Dashed gray lines are included as guides for the eye and were obtained by the extrapolation of 
 the ED results for $L\leq 14$. 
 } 
 \end{figure}
Elementary excitations further reveal the topological nature of the Pfaffian-like state. By construction, the Pfaffian-like ansatz, (\ref{eq:Pfaffian1D_2}), has a hidden global order associated with the organization of particles in two 
identical indistinguishable copies of the same state. Consequently, the elementary excitations above the Pfaffian-like ground-state ansatz exhibit non-Abelian statistics \cite{Paredes3}. 
The argument for this proceeds as follows. The elementary excitations can be constructed by creating a quasihole in each of the copies and symmetrizing \cite{Paredes3}. 
Symmetrization leads to a topological degeneracy in the subspace of elementary excitations and non-Abelian algebra of exchanges of elementary excitations 
(quasiholes) \cite{Paredes3}.

If the ground state of the system is close to the Pfaffian-like ansatz, elementary excitations above it also exhibit non-Abelian statistics. 
To confirm this statement we compare excited states of the Hamiltonian, (\ref{eq:Hamiltonian5}), with the ansatz wave functions for the 
excited states \cite{Paredes3}
\begin{equation}\label{eq:ansatz_es}
|\psi_3^{(m,n)}\rangle=\mathcal{P}(|\Psi_2^{\uparrow(m)}\rangle \otimes |\Psi_2^{\downarrow(n)}\rangle),
\end{equation}
where $|\psi_2^{\sigma(k)}\rangle$ are eigenstates for each of two copies. In particular, we calculate the overlap of the first excited state of the Hamiltonian, (\ref{eq:Hamiltonian5}),
and corresponding ansatz wave function for the first excited state, (\ref{eq:ansatz_es}).

For a fixed number of particles $N$ (with $N$ either even or odd) an excited state 
with $2n$ elementary excitations (quasiholes) that are SU(2)$_2$ (Ising) anyons (similar to Majorana fermions) is expected to have topological degeneracy $2^{n-1}$ \cite{Greiter}.
We find that the first excited state of the Hamiltonian, (\ref{eq:Hamiltonian5}), is twofold degenerate and corresponds to a state with four quasiholes.

The overlap for each of the two degenerate first excited states is calculated by considering the total overlap with the manifold of degenerate ansatz states, (\ref{eq:ansatz_es}), 
that correspond to the first excited states. We find four degenerate, linearly independent, ansatz states that correspond to two degenerate first excited states of the 
Hamiltonian, (\ref{eq:Hamiltonian5}). The degeneracy of the exact first excited state is two, and not four, since the Hamiltonian, (\ref{eq:Hamiltonian5}), does not have particle-hole
 symmetry. The overlap is then given by
 \begin{equation}\label{eq:ovl_es}
 O_i=\sqrt{\sum_k|\langle \psi^{(1)}_i|\phi_k^{(1)} \rangle|^2}\equiv |\langle \psi^{(1)}_{Exact} | \psi^{(1)}_{Ansatz}\rangle|_i,
 \end{equation}
 where $|\psi^{(1)}_i\rangle$, with $i=1,2$, are two degenerate first excited states of the Hamiltonian, (\ref{eq:Hamiltonian5}), and $|\phi^{(1)}_k\rangle$, with $k=1,...,4$,
 are corresponding degenerate ansatz states. We note that the states $|\phi^{(1)}_k\rangle$ form an orthonormal basis within the degenerate manifold, which leads to expression 
 (\ref{eq:ovl_es}) for the total overlap. We find that $O_1=O_2$ for all values of the two-body interaction strength $U/t$ that we have considered.

The overlap of the first excited states and corresponding ansatz wave functions is shown in Fig. \ref{fig:Overlap0Uc_es}. In agreement with the results for the ground-state 
wave function, the overlap decreases more gradually with increasing system size $L$ at $\bar{U}_C(L)$ than it decreases at $U=0$ (Fig. \ref{fig:Overlap0Uc}). 
The results also demonstrate that the overlap with the ansatz wave function for four quasiholes is very close to 1,  indicating that the elementary excitations 
of the system are non-Abelian. 

\section{Variational Monte Carlo calculation}
\label{sec:VMC}
\begin{figure}[b!]
\includegraphics[width=\columnwidth]{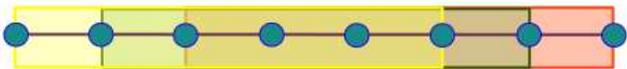}
 \caption{\label{fig:Plaquettes} Illustration of the six-site 1D plaquettes with a five-site overlap between the plaquettes used in the calculation of the ground-state
 properties of the system for larger system sizes.} 
 \end{figure}
Studying properties of the system for larger system sizes with the ED method is not possible due to the rapid increase in the Hilbert-space size with increasing  
system size. Motivated by successes of the tensor network methods \cite{Orus} to numerically simulate a variety of strongly correlated models, we further study the properties of
the system for larger system sizes using the entangled-plaquette-state ansatz optimized using the variational Monte Carlo method \cite{Changlani,Mezzacapo1,Neuscamman1,Mezzacapo2,Mezzacapo3,Mezzacapo4,Mezzacapo5,Neuscamman2,AlAssam,Duric}. 
Within the EPS approach, also called the correlator-product-state approach, the lattice is covered with overlapping plaquettes and the ground-state wave function is written in terms of the plaquette coefficients. 
Configurational weights can then be optimized using a VMC algorithm.

For a lattice with $L$ sites, an arbitrary quantum many-body wave function can be written as
\begin{equation}
|\psi\rangle=\sum_{n_1,...,n_L}W_{n_1,...,n_L}|n_1,...,n_L\rangle=\sum_{\textbf{n}}W_{\textbf{n}}|\textbf{n}\rangle,
\end{equation}
where $\textbf{n}=\left\{n_1,...,n_L\right\}$ denotes the vector of occupancies and $W_{\textbf{n}}$ is the amplitude or weight of a given configuration $\textbf{n}$. For the system described by the 
Hamiltonian (\ref{eq:Hamiltonian5}) $n_i\in\left\{0,1,2\right\}$ for the lattice sites $i=1,...,L$.
\\
\indent
In the EPS (CPS) description of a bosonic system, the weight $W_{\textbf{n}}$ is expressed as a product of the plaquette coefficients over the lattice 
\begin{equation}\label{eq:Wn}
W_{n_1,...,n_L}=\prod_p C_p^{\textbf{n}_p},
\end{equation}
where $\textbf{n}_p=\left\{n_{p1},...,n_{pl}\right\}$ is the occupancy vector of the $l$-site plaquette $p$. In many cases, the qualitative behavior of large systems can be 
described even by plaquettes with a small number of sites. The EPS wave function corresponding to the ground-state 
wave function of the system gives reasonable estimates of the ground-state energy and short-range correlations. 
The estimates improve with an increase in plaquette size and greater overlap between the plaquettes.

Here we choose six-site 1D plaquettes with a five-site overlap 
between the plaquettes as illustrated in Fig. \ref{fig:Plaquettes}. The weights $W_\textbf{n}$ for the ground-state wave function of Hamiltonian (\ref{eq:Hamiltonian5}) with periodic boundary conditions 
can then be written as
\begin{equation}\label{eq:Wn_6}
W_{\textbf{n}}=C_1^{n_1,n_2,...,n_6}\cdot C_2^{n_2,n_3,...,n_7}\cdot ... \cdot C_L^{n_L,n_1,...n_5}. 
\end{equation}
\begin{table}[t!]
\centering
\begin{tabular}{c c c c c c c}

$U/t$ & $E_{0,EPS}$  & $E_{0,ED}$ & R($\times 10^{2}$)\\
\hline
0.0&-1.61170&-1.62516&0.828\\
0.5&-1.49408&-1.50630&0.811\\
1.0&-1.38258&-1.39293&0.743\\
$\approx 1.18 \approx \bar{U}_C$&-1.33919&-1.35329&1.042&\\
1.5&-1.28014&-1.28519&0.393\\
2.0&-1.17054&-1.18326&1.075\\
2.5&-1.07458&-1.08740&1.179\\
3.0&-0.98504&-0.997966&1.295\\
\hline
\end{tabular}
\caption{Ground-state energy per site (in units of $t$) for the system size $L=14$ and with periodic boundary conditions. Here $R$ stands for 
the relative error with respect to the exact ground-state energy and is defined as $R=(E_{0,ED}-E_{0,EPS})/E_{0,ED}$.}
\label{table:E0}
\end{table}
Within the variational MPS approach\cite{Paredes1} this plaquette choice would correspond to matrices of dimension $\chi=3^{L^o_p}$, with $L_p^o=5$ being 
the number of overlapping lattice sites.

 In a VMC algorithm the energy $E$ is written as
 \begin{eqnarray}\label{eq:E_VMC}
 E&=&\frac{\langle\psi|H|\psi\rangle}{\langle\psi|\psi\rangle}=\frac{\sum_{\textbf{n},\textbf{n}'}W^*_{\textbf{n}'}\langle\textbf{n}'|H|\textbf{n}\rangle W_{\textbf{n}}}{\sum_{\textbf{n}}|W_{\textbf{n}}|^2}\nonumber\\
 &=&\sum_{\textbf{n}}P_\textbf{n}E_\textbf{n},
 \end{eqnarray}
 where it is assumed that the wave function $|\psi\rangle=\sum_{\textbf{n}}W_\textbf{n}|\textbf{n}\rangle$ is not normalized, and the local energy $E_{\textbf{n}}$ and the 
 probability $P_{\textbf{n}}$ are given by
 \begin{eqnarray}\label{eq:E_VMC_2}
 E_{\textbf{n}}&=&\sum_{\textbf{n}'}\frac{W_{\textbf{n}'}^*}{W_{\textbf{n}}^*}\langle \textbf{n}'|H|\textbf{n}\rangle,\\
 P_{\textbf{n}}&=&\frac{|W_\textbf{n}|^2}{\sum_{\textbf{n}}|W_\textbf{n}|^2}.\nonumber
 \end{eqnarray}
 The expectation value of any operator $\hat{O}$ can be expressed in the same form by replacing the Hamiltonian $H$ with the operator $\hat{O}$. The probability $P_\textbf{n}$
  is never explicitly calculated from Eq. (\ref{eq:E_VMC_2}). Instead, for a given set of plaquette coefficients, the energy can be efficiently computed using the Metropolis 
  algorithm \cite{Metropolis}.

  \begin{figure}[t!]
  \includegraphics[width=\columnwidth]{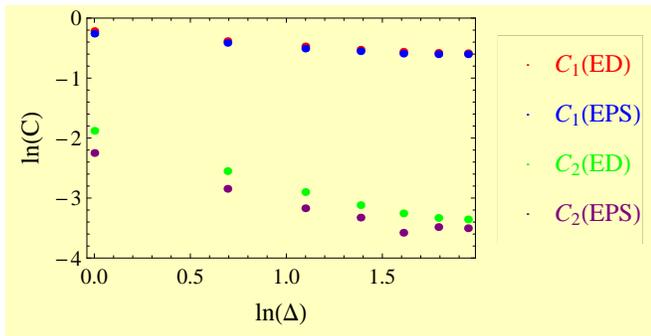}
 \caption{\label{fig:CF_12_L14}The one-body (red and blue symbols) and two-body (green and purple symbols) correlation functions obtained using the ED (red and green symbols) and the EPS and VMC 
 (blue and purple symbols) methods for the system size $L=14$ and at the value of the two-body interaction strength $U/t\approx \bar{U}_C(L)\approx 1.18$.} 
 \end{figure}
  Within the Metropolis algorithm, used to sample the probability distribution, the overall energy can be efficiently computed as an average of the sampled local energies. 
  In our calculation the total number of atoms $N$ is fixed. We start from a randomly chosen initial configuration $|\textbf{n}\rangle=|n_1,n_2,...,n_L\rangle$, with 
  $\sum_{i=1}^L n_i=N$ and $n_i\in \left\{0,1,2\right\}$ for $i=1,...,L$, and then generate via the Metropolis algorithm a large set of new configurations by replacing 
  $n_i$ with $n_i-1$ (if $n_i>0$) and $n_j$ with $n_j+1$ (if $n_j<2$) at two neighboring sites $i$ and $j$. Starting from configuration $|\textbf{n}\rangle$, the 
  acceptance probability of a new configuration $|\textbf{n}'\rangle$ is given by
  \begin{equation}\label{eq:P_A}
  P_A=\text{min} \left[\frac{|W(\textbf{n}')|^2}{|W(\textbf{n})|^2},1\right].
  \end{equation}

  According to the variational principle, minimization of expression (\ref{eq:E_VMC}) with respect to the weights gives an upper bound of the ground-state energy. 
  The plaquette coefficients that minimize the energy can be found by using the stochastic minimization method \cite{AlAssam,Duric,Metropolis,Foulkes, Sandvik, Lou}, which 
  requires only the first derivative of the energy with respect to the plaquette coefficients, which is given by 
  \begin{equation}\label{eq:D_E}
  \frac{\partial E}{\partial C_p^{\textbf{n}_p}}= 2\sum_{\textbf{n}}\left\{ P_\textbf{n}\Delta_p^{\textbf{n}_p}\left[ E_\textbf{n}-\sum_{\textbf{n}'}P_{\textbf{n}'}E_{\textbf{n}'}\right]\right\},
  \end{equation}
 where the wave function $|\psi\rangle$ in Eq. (\ref{eq:E_VMC}) is approximated by the EPS (CPS) wave function and 
 \begin{equation}\label{eq:Delta_p}
 \Delta_p^{\textbf{n}_p}=\frac{1}{W_{\textbf{n}}}\frac{\partial W_{\textbf{n}}}{\partial C_p^{\textbf{n}_p}}=\frac{b_p}{C_p^{\textbf{n}_p}},
 \end{equation}
 with $W_{\textbf{n}}$ given by Eq. (\ref{eq:Wn_6}). Here $b_p$ denotes the number of times the plaquette coefficient $C_p^{\textbf{n}_p}$ appears in the product, (\ref{eq:Wn_6}), for 
 the amplitude $W_{\textbf{n}}$ for configuration $|\textbf{n}\rangle$. If the same plaquette coefficient is used for multiple sites (e.g., for a translationally invariant choice of plaquettes), 
 $b_p>1$. If each plaquette coefficient is used once, $b_p=1$. Equivalently to the overall energy the first derivative can be efficiently 
 calculated using the Metropolis algorithm from the same sample used to compute the overall energy.

\begin{figure}[b!]
\includegraphics[width=\columnwidth]{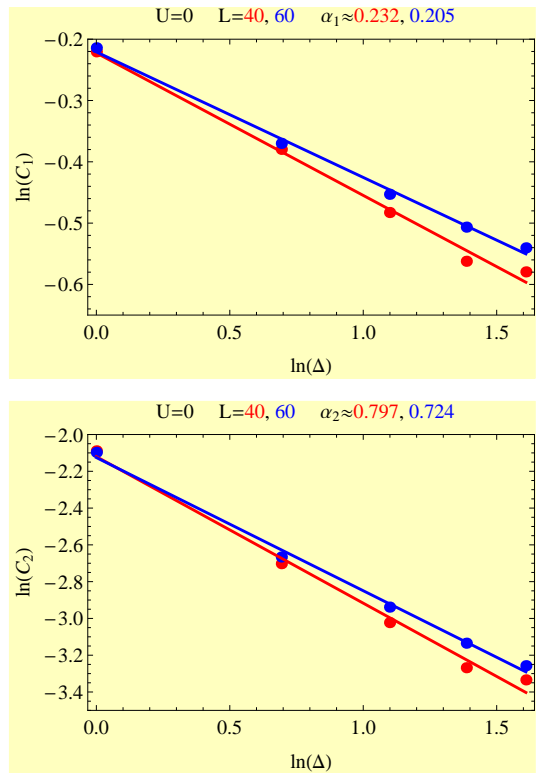}
 \caption{\label{fig:CF_U0} The one-body ($C_1$) and two-body ($C_2$) correlation functions for the system sizes $L=40$ (red symbols), and $L=60$ (blue symbols) sites and at the
 value of the two-body interaction strength $U/t=0$. 
 Here the asymptotic behavior of the correlation functions is $C_1=\langle a_{i+\Delta}^\dagger a_i\rangle\rightarrow \Delta^{-\alpha_1}$ 
 and $C_2=\langle a_{i+\Delta}^\dagger a_{i+\Delta}^\dagger a_i a_i\rangle\rightarrow \Delta^{-\alpha_2}$. } 
 \end{figure}
 The steps of the VMC algorithm used to calculate the ground-state properties of the system are as follows: (i) start from the randomly chosen complex values for the 
 plaquette coefficients, (ii) evaluate the energy and its gradient vector, (iii) update all plaquette coefficients $C_p^{\textbf{n}_p}$ according to
 \begin{equation}
 C_p^{\textbf{n}_p}\rightarrow C_p^{\textbf{n}_p}-r\delta(k)\cdot\text{sign}\left(\frac{\partial E}{\partial C_p^{\textbf{n}_p}}\right)^*,
 \end{equation}
 and (iv) iterate from (ii) until convergence of the energy is reached. Here $r$ is a random number between $0$ and $1$, and $\delta(k)$ is the step size for a given iteration $k$.

 \begin{figure}[t!]
 \includegraphics[width=\columnwidth]{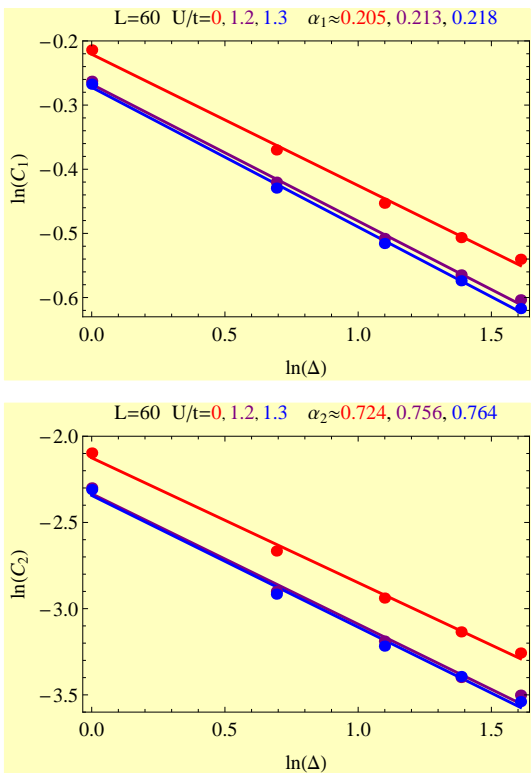}
 \caption{\label{fig:CF_L60} The one-body ($C_1$) and two-body ($C_2$) correlation functions for the system size $L=60$ sites and at the values of the two-body interaction strength $U/t=0$ (red symbols), 
 $1.2$ (purple symbols), and $1.3$ (blue symbols). 
 Here the asymptotic behavior of the correlation functions is $C_1=\langle a_{i+\Delta}^\dagger a_i\rangle\rightarrow \Delta^{-\alpha_1}$ and 
 $C_2=\langle a_{i+\Delta}^\dagger a_{i+\Delta}^\dagger a_i a_i\rangle\rightarrow \Delta^{-\alpha_2}$.} 
 \end{figure}
 In each iteration $k$, the energy and its derivative are estimated from $F(k)\times L$ values, where $L$ is the number of lattice sites and $F(k)$ is called the number of 
 sweeps per sample. In a given sweep each lattice site is visited sequentially and a move, $\textbf{n}\rightarrow\textbf{n}'$, to a new configuration is proposed by changing the 
 occupancy numbers $n_i\rightarrow n_i-1$ (for $n_i>0$) and $n_j\rightarrow n_j+1$ (for $n_j<2$) at two neighboring lattice sites, $i$ and $j$. Also, to achieve convergence and 
 reach the optimal energy value it is important to carefully tune the gradient step $\delta(k)$. For each iteration $k$, the number of sweeps $F$ is increased linearly, $F = F_0k$, 
 and the procedure of evaluating the energy and updating the coefficients is repeated $G = G_0k$ times. The step size is gradually reduced per iteration. Here we use 
 a geometric form, $\delta= \delta_0Q^k$ , with $Q = 0.9$.

The number of sweeps per iteration is increased because the derivatives become smaller as the energy minimum is approached and require more sampling in order not to be
dominated by noise. An increasing $G$ effectively corresponds to a slower cooling rate. Here we take $F_0 = 100$, $G_0 = 10$, and $Q = 0.9$. The initial minimization 
routine is performed with $\delta_0 = 0.5$ for 50 iterations. The resulting plaquette coefficients are then used as a starting point for a new run of 50 iterations
with $\delta_0 = 0.05$. After the minimization is complete the expectation values are calculated by repeating the procedure for a single iteration with 0 step size and 
large $F$ and $G$ to obtain more accurate estimates of the expectation values.

It is also important to note that it is more difficult to obtain good estimates of the ground-state energies and the correlation functions for small system sizes due to the 
presence of the statistical error in the stochastic algorithm. Having a larger number of parameters allows the optimization method more freedom in finding the minimum energy 
state and the statistical error can be controlled by increasing the system size.

The results for the ground-state energy for the system size $L=14$ and for several values of the two-body interaction strength $U/t$ are listed in Tab. \ref{table:E0} and compared 
to the ED results (exact ground-state energy values). As can be seen in Table \ref{table:E0}, the relative error, defined as
\begin{equation}\label{eq:R}
R=\frac{E_{0,ED}-E_{0,EPS}}{E_{0,ED}}, 
\end{equation}
does not exceed $1.3\%$ for any value of the two-body interaction strength $U/t$.
The EPS and VMC calculation also gives quite accurate estimates of the correlation functions as demonstrated in Fig. \ref{fig:CF_12_L14}.

To examine the proximity of the ground-state wave function for larger system sizes to the Pfaffian-like ansatz wave function, we further calculate correlation functions 
for system sizes $L=40$ and 60 sites. Since the EPS wave function gives quite accurate estimates of the correlations within any plaquette $p$, to estimate 
the asymptotic behavior of the correlation functions we calculate the one-body and two-body correlation functions, (\ref{eq:C1C2}), for the lattice sites within a plaquette 
$p$ ($\Delta=1,...,l_p-1$).

The results for the one-body and two-body correlation functions at $U=0$ and for the system sizes $L=40$ and 60 sites are shown in Fig. \ref{fig:CF_U0}. The values of 
 $\alpha_1$ and $\alpha_2$ that describe the asymptotic behavior of the correlation functions,
\begin{eqnarray}\label{eq:CF_12}
C_1&=&\langle a_{i+\Delta}^\dagger a_i\rangle\rightarrow \Delta^{-\alpha_1},\\
C_2&=&\langle a_{i+\Delta}^\dagger a_{i+\Delta}^\dagger a_i a_i\rangle\rightarrow \Delta^{-\alpha_2},\nonumber
\end{eqnarray}
move away from the values expected for the Pfaffian-like state ansatz ($\alpha_1=0.25$ and $\alpha_2=1$) with increasing system size. The results thus suggest that the overlap between 
the exact ground-state wave function at $U=0$ and the Pfaffian-like ansatz decreases with increasing system size. This is consistent with the ED results 
and with the previously obtained results for the system sizes $L\leq40$ obtained using the variational MPS \cite{Paredes1}.

The ED results presented in the previous section also suggest that the Pfaffian-like ansatz wave function, (\ref{eq:Pfaffian1D_2}), better approximates the 
exact ground state of the Hamiltonian, (\ref{eq:Hamiltonian5}), at some finite value of the two-body interaction strength $U/t=\bar{U}_C(L)$ than it approximates 
the exact ground-state wave function at $U=0$. The results for the one-body and two-body correlation functions for the system size $L=60$ sites (Fig. \ref{fig:CF_L60}) show that the values of $\alpha_1$ and $\alpha_2$ increase with an increase in the value of the two-body interaction strength $U/t=\bar{U}$.
In other words, at some value $\bar{U}_C(L)$ the values of $\alpha_1$ and $\alpha_2$ will be the closest to the values expected for the Pfaffian-like ansatz wave function 
($\alpha_1=0.25$ and $\alpha_2=1$). This indicates that the overlap between the exact ground-state wave function and the Pfaffian-like ansatz wave function is maximal at 
$\bar{U}_C(L)$, in agreement with the ED results for smaller system sizes.

Previous calculations with variational MPS found the values of $\alpha_1$ and $\alpha_2$ for the system size $L=20$ sites to be $\alpha_1=0.22$ and $\alpha_2=0.83$
 for the exact ground-state wave function at $U=0$ and $\alpha_1=0.24$ and $\alpha_2=0.99$ for the Pfaffian-like ansatz wave function \cite{Paredes1}. The corresponding overlap between the 
 exact ground-state and the Pfaffian-like state wave functions was found to be $\approx 0.955$ \cite{Paredes1}. Also, for the system size $L=40$ sites at $U=0$ the overlap is
 $\approx 0.90$ \cite{Paredes1}, which corresponds to $\alpha_1\approx0.232$ and $\alpha_2\approx0.797$ obtained within our EPS and VMC calculation for the exact ground-state 
 wave function (Fig. \ref{fig:CF_U0}) and $\alpha_1\approx0.25$ and $\alpha_2\approx 1$ for the Pfaffian-like ansatz wave function.

It is difficult to determine the exact values of $\bar{U}_C(L)$ from our EPS and VMC calculation. However, for the system size $L=60$ sites, the maximum system size that 
we have considered, we find that $\alpha_1\gtrsim 0.218$ and $\alpha_2\gtrsim 0.764$ at $U/t=\bar{U}_C(L)$. Therefore, based on the results mentioned in the previous paragraph, we 
estimate that the overlap between the exact ground-state and the Pfaffian-like-state wave functions is still good for the system size $L=60$ and at $U/t=\bar{U}_C(L)$. 
\section{Conclusions}
\label{sec:Conclusions}
We have studied ground states and elementary excitations of a system of bosonic atoms and diatomic Feshbach molecules trapped in a 1D optical lattice. Under certain conditions, which are experimentally 
achievable with current technology in systems of cold atoms and molecules in optical lattices, the system can be described by an effective Hamiltonian for bosonic atoms with 
two- and three-body interactions. We have considered the limit of infinitely strong three-body interactions for a range of values of the two-body interaction strength.
The ground-state properties of the system were calculated using the ED method for small system sizes, and the EPS and VMC method for larger system sizes.

The Pfaffian-like ansatz was originally proposed as an ansatz for the ground-state wave function of the effective Hamiltonian in the absence of two-body interactions. However, our results 
clearly demonstrated that the Pfaffian-like ansatz wave function is a better ansatz for the ground-state wave function of the effective Hamiltonian at some finite value 
of the two-body interaction strength. This value of the two-body interaction strength might be close to the value where the system undergoes a quantum phase transition 
from the superfluid state to the Mott insulating state, as previously found within the bosonization approach. We also demonstrate that these states support non-Abelian excitations 
required for quantum computation.

Further work is necessary to find an experimentally realizable model with a ground-state wave function that can be even better approximated by the Pfaffian-like ansatz 
wave function. This can possibly be achieved in a system with long-range interactions. An additional direction for future research is to consider similar 2D non-Abelian models, for 
example, anisotropic systems consisting of coupled interacting 1D wires. Such anisotropic 2D lattice models can have interesting non-Abelian Chern insulating phases and fractional topological 
insulating phases.

In order to use non-Abelian states for quantum computation, one must be able to braid them. Although this is not possible for a strictly 1D system, creating a network of such 
1D systems connected by T-junctions, as suggested previously in the context of Majorana quantum wires \cite{Alicea}, potentially allows this. In the case of Majorana quantum wires a 
T-junction \cite{Alicea} allows for adiabatic exchange of two Majorana fermions. Such a T-junction has topological and non-topological regions that can be controlled by individually 
tunable gates. In principle, similar T-junction networks can be created for the bosonic system that we have considered, where topological and nontopological regions of the network can be
controlled by tuning the two-body interaction in different regions of the T-junction network (for example, by changing the depth of the optical lattice in certain regions of the T-junction).

In a similar manner to fractional quantum Hall states, non-Abelian anyons can be created by creating pairs of quasiholes, with one quasihole in each cluster \cite{Paredes1,Paredes3}. These non-Abelian anyons 
are Ising anyons [SU(2)$_2$ anyons], similar to Majorana fermions, and excitations of the  $\nu = 5/2$ fractional quantum Hall state (Pfaffian state). We also note that braiding of SU(2)$_2$ anyons alone 
does not permit universal quantum computation \cite{Nayak}. However, to obtain a universal set of gates, braiding of SU(2)$_2$  anyons needs to be strengthened only by a single-qubit 
$\pi/8$  phase gate and a two-qubit measurement \cite{Nayak}. Also, Pfaffian-like states obtained by symmetrization of two identical copies can be generalized to states obtained by 
symmetrization of $k$  identical copies that support SU(2)$_k$ anyons \cite{Paredes2,Paredes3} and can be used for universal quantum computation. For example SU(2)$_3$ anyons 
(like Fibonacci anyons) can be used for universal quantum computation \cite{Nayak}. The results presented here thus constitute an important step towards understanding generalized states 
that support SU(2)$_k$ anyons. 
\begin{acknowledgments}
We thank Jakub Zakrzewski for very helpful suggestions and discussions. 
We acknowledge support from the EPSRC through Grants Nos. EP/K02163X/1, and EP/I004831/2, and TOPNES
program Grant No. EP/I031014/1. N. Chancellor was funded by Lockheed Martin Corporation at the time this work was carried out. T. \DJ uri\'c  also acknowledges support from the EU Grant QUIC (H2020-FETPROACT-2014, Grant No. 641122).
\end{acknowledgments}

\end{document}